\documentclass[aps,pra,floatfix,showpacs,twocolumn,superscriptaddress]{revtex4}
\usepackage{amsmath}
\usepackage{graphicx}

\newcommand{\ket}[1]{\mbox{$|#1\rangle$}}

\newcommand{\Ket}[1]{\mbox{$|\textbf{#1}\rangle$}}

\newcommand{\ave}[1]{\langle #1 \rangle}

\begin{document}

\title{Utilizing Encoding in Scalable Linear Optics Quantum Computing}

\author{A.~J.~F.~Hayes}

\address{Centre for Quantum
Computer Technology and Physics Department, University of Queensland, QLD 4072, Brisbane,
Australia.}

\author{A.~Gilchrist}
\email{alexei@physics.uq.edu.au}
\address{Centre for Quantum
Computer Technology and Physics Department, University of Queensland, QLD 4072, Brisbane,
Australia.}

\author{C.~R.~Myers}
\address{Institute for Quantum Computing, University of
Waterloo, ON, N2L 3G1, Canada}

\author{T.~C.~Ralph}
\address{Centre for Quantum
Computer Technology and Physics Department, University of Queensland, QLD 4072, Brisbane,
Australia.}


\begin{abstract}
We present a scheme which offers a significant reduction in the resources
required to implement linear optics quantum computing.  The scheme is a
variation of the proposal of Knill, Laflamme, and Milburn, and makes use of an
incremental approach to the error encoding to boost probability of success. 
\end{abstract}


\maketitle
\section{Introduction}

%
%
There are many different physical systems being considered for  a quantum
computer. Of these, optical systems are promising, offering benefits such as
low decoherence and easy single qubit manipulation.  Multi qubit gates are
beginning to be demonstrated \cite{03pittman032316,03obrien264,gasparoni,zhao}
and the next stage is to start linking several gates together into small scale
algorithms.  Finding efficient ways of implementing these gates will be crucial
to the overall success of the program.  In this paper we show that the the
resources required to implement linear optics quantum computation (LOQC) based
on the ideas in KLM \cite{klm} can be significantly reduced.

%
%
A universal set of gates is necessary for quantum computing to be possible, and
in LOQC a convenient set to choose is an entangling gate such as a
controlled-not (\textsc{cnot}) together with arbitrary single qubit gates. To
be able to generate an arbitrary transformation on a single qubit, it suffices
to produce arbitrary rotations in one direction on the Bloch sphere, and $90$
degree rotations in an orthogonal direction.

%
%
In LOQC the qubit is typically encoded in ``dual-rail'' form
\footnote{In dual-rail encoding the qubit states could be formed from the
polarisation states of a photon.  The polarisation modes are morphologically
equivalent to using two spatial modes, where the presence of the photon in one
mode represents a logical ``0'' and in the other a logical ``1''. Either
spatial or polarisation modes may be chosen depending on the practical
considerations of the circuit.  The alternative, ``single-rail'' encoding, uses
the presence or absence of a photon in a single mode to represent the state of
the logical qubit.}.
With dual-rail encoding, arbitrary deterministic rotations in both the $X$
and $Z$ directions\footnote{We shall denote a qubit operation that applies a
rotation around $X$ by $\theta$ as $X_\theta$} are possible with only passive linear
optical elements (such as phase shifters and beam-splitters) meaning that any
single qubit transformation can be performed with relative ease.
%
%
The real difficulty arises in performing a \textsc{cnot} gate between single
photon qubits.  This requires a nonlinearity sufficiently large that a single
photon can induce a $\pi$ phase shift on another, and such a large intrinsic
nonlinearity is not possible with current materials.

%
%
In 2001, Knill, Laflamme \& Milburn (KLM) \cite{klm} presented a way of
implementing scalable quantum computing using single photons, linear optics and
photo-detection.  Vital to the proposal was the elimination of the need for the
non-linear coupling between photons, replacing it instead with a heralded
nondeterministic \textsc{cnot} gate (that is, a \textsc{cnot} gate that does not
always work; but when it does, it is known to have worked). The gate KLM described functions
with a success probability of $1/16$.  Later work resulted in a number of
variations on the basic non-deterministic gate \cite{ralphgate, knillgate,01pittman062311}.

%
%
Using a non-deterministic \textsc{cnot} gate is only the first step, as clearly,
chaining many of them together cannot be efficiently scalable.  To allow
non-deterministic gates to be performed on the qubits with a high probability
of success, KLM suggested using teleportation \cite{gottnike}. This will allow the
gates to be performed on an entangled state, which is
then used as resource for the circuit. By using this modified entanglement to
teleport the input qubit, the gate will be applied to the qubit with certainty
in the event of a successful teleportation.

%
%
A key complication in LOQC  schemes is a lack of deterministic Bell
measurement, hence the teleportation itself cannot be performed
deterministically.  The KLM proposal counteracts this with a family of
teleporters which asymptotically approach a success rate of 100\% as the
complexity of the required entanglement increases. In addition, a teleportation
failure looks like a $Z$-measurement error (an accidental measurement in the
computational basis), and error correction can be used to protect against this
--- for a recent experimental demonstration of the technique see
\cite{zmeasexp}.

%
%
In this paper we propose using an incremental approach to the encoding. This
approach simplifies the process of gate attempts and recovery, and leads to
improved procedures for performing gates on the encoded qubits.  Issues of
fault tolerance are not covered in this paper; the techniques described here
deal only with the intrinsic errors resulting from the use of non-deterministic
systems.  The aim is to find efficient implementations of the gates for use in
LOQC.

%
%
Section~\ref{sec:klm} of this paper contains a brief overview of the salient concepts
and procedures used in the original KLM proposal. Section~\ref{sec:incremental}
presents our alternative approach to producing and operating on encoded qubits.
Section~\ref{sec:probs} contains an analysis of the probabilities of operation
of the encoded gates, and Section~\ref{sec:resources} tallies the resources
required by our incremental scheme.

%
%
Recently some publications have considered promising
implementations of linear optics quantum computing using cluster
state architectures produced via KLM-style gates \cite{yorrez,
cluster,browne}. It should be noted that the scheme in this paper has similarities to
the cluster state models, since the process of encoding and
re-encoding is closely related to the process used to build up a
cluster state. However, our discussion is restricted to the
standard quantum circuit type architecture.

\section{Review of KLM}
\label{sec:klm}

We briefly review the key features of the KLM proposal.

\subsection{Hiding Under Teleportation}

%
%
The KLM proposal describes a family of teleporters T$_{n/n+1}$ (where $n$ is a
positive integer) which function with a probability $n/(n+1)$, each needing a
successively larger entanglement resource to be prepared.   The general
entanglement resource for teleportation is:
\begin{equation}
\ket{t_n}=\frac{1}{\sqrt{n+1}}\sum_{j=0}^n
        \ket{1}^j\ket{0}^{n-j} \ket{0}^j\ket{1}^{n-j}.
\label{tnform1}
\end{equation}
Note that the most basic teleporter has a success probability of $1/2$ and
its resource state can be prepared with just beam-splitters.

\subsection{$Z$-Measurement Error Correction}

%
%
Besides improving the probability of successfully performing a
gate, the teleportation scheme has the advantage of a well-defined
failure behaviour. In the event that the teleportation fails, it
acts as a $Z$-measurement with known outcome on the qubit which was
to be teleported. Hence the probability of a successful
teleportation can be improved by using error correction to protect
the qubits from $Z$-measurements
\footnote{
%
%
A variation on the KLM teleporters with a higher average fidelity was proposed
by Franson, \emph{et al.} \cite{frans}. It has an error rate of $1/n^2$, but is
not useful in this proposal since it lacks the well-defined failure mode
required by the scheme.}.

%
%
Encoding against unwanted $Z$-measurements is one of the simplest
forms of error correction. A two-qubit encoding is given as an
example in the KLM proposal (note that kets containing bold font are used to
indicate logical qubit states, as distinct from mode occupation numbers):
\begin{equation}\label{ZMcode}
\begin{split}
   \Ket{0}\rightarrow&\frac{1}{\sqrt{2}}(\Ket{00}+\Ket{11})\\
   \Ket{1}\rightarrow&\frac{1}{\sqrt{2}}(\Ket{01}+\Ket{10}),
\end{split}    
\end{equation}
so that an arbitrary qubit $\Ket{$\Psi$}=\alpha\Ket{0}+\beta\Ket{1}$ becomes:
\begin{multline}
\Ket{$\Psi$}\rightarrow\Ket{$\Psi$}^{(2)}=
\frac{1}{\sqrt{2}}[\alpha(
\Ket{00}_{AB}+\Ket{11}_{AB})+\\
\beta(\Ket{01}_{AB}+\Ket{10}_{AB})].
\end{multline}
The notation $\Ket{$\Psi$}^{(2)}$ represents a state encoded
across two qubits.

%
%
Here the logical qubit can be recovered if a single
$Z$-measurement occurs, as long as the result of the measurement and
which qubit it occurred on are known. The effect of a measurement
of ``0'' on one of the qubits is to leave the other qubit in the
original unencoded state. In the event that a ``1'' is measured, an
$X$ gate must first be performed on the remaining qubit in order to
reproduce the original state. Therefore, correcting an unwanted
measurement will require a possible $X$ operation, followed in both cases by re-encoding
the state to $\Ket{$\Psi$}^{(2)}$.

\subsection{Concatenation Encoding}

%
%
To apply the encoding across more than two qubits, KLM uses a concatenated
approach where the same encoding and operational procedures are applied to each
level of the encoded qubits.
For example,
\begin{multline}
\Ket{$\Psi$}^{(4)}=\frac{\alpha}{\sqrt{2}}
(\Ket{00}^{(2)}+\Ket{11}^{(2)})\\+\frac{\beta}{\sqrt{2}}(\Ket{01}^{(2)}+
\Ket{10}^{(2)}).
\end{multline}
Each concatenation doubles the number of component
qubits used, but the total probability of failure on a gate performed at that
level decreases. As a result, the chance of failure can be made arbitrarily
small by successive concatenations of the code. Of course, the 
resources required to perform the
encoding also increases.

%
%
A variation of the teleporter circuit is described in KLM for use
in the higher levels of encoding (Figure 5a of \cite{klm}). The entanglement resource used
to create this teleporter is denoted \ket{tx_n}. This circuit is
formed using the fundamental set of gates described earlier, and
will teleport a qubit with certainty if the gates are performed
successfully. If the gates in the teleporter fail, the overall
effect is again a $Z$-measurement on the qubit to be teleported. Hence the
problem of performing gates on higher levels reduces to one of
successful teleportation at the lowest level.

\begin{figure}[htb]
\centering
\includegraphics[height=180pt,width=180pt]{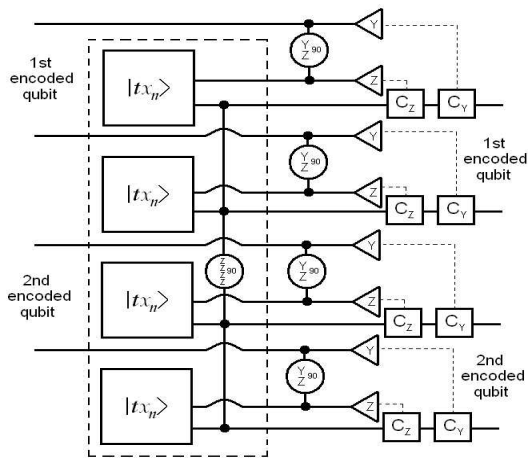}
\caption{Performing a logical \textsc{cs} in the KLM scheme. C$_Z$ and
C$_Y$ indicate correction gates which depend on the measurements
made when teleporting, and are described in the KLM paper. The circled gates
indicate an operation of the form $\exp\{-i\pi Y\otimes Z/4\}$ and are also
described in the KLM paper. The dashed box indicates the resource state for
the gate, and this can be prepared off-line.
}
\label{figKLMG}
\end{figure}

%
%
Figure \ref{figKLMG} shows the manner in which an encoded controlled-phase-flip
gate (\textsc{cs} gate) is constructed in the KLM scheme. Each \textsc{cs} gate
performed at a particular level of encoding is implemented by applying this
circuit to the next level down.

\subsection{Success Probability For an Encoded Gate}

%
%
There are two non-deterministic operations in the set required by KLM, when
acting on encoded states. These are the $Z_{90}$ and \textsc{cs} gates. KLM
suggest circuits for implementing these gates, to be applied iteratively at
each level of concatenation. These circuits use teleportation networks built
from the basic set of gates, and accordingly each requires its own resource
state, which must also be built up iteratively.

%
%
The possibility of measurement error recovery is considered in the failure
probability given for the two non-deterministic gates.  When the code is
concatenated, increasing the size of the encoding, the failure probabilities
for these gates drop correspondingly.  The failure probability for teleporting
a logical qubit, given in \cite{thresholds}, is:
\begin{equation}
F_Z=f^2(2-f)/(1-f(1-f)),
\end{equation}
where $f$ is the probability that a teleportation on one of the two component
qubits fails. $F_Z$ is the probability for a encoded $Z_{90}$ rotation failing,
which requires one teleportation of a logical qubit. The probability of
successfully applying a \textsc{cs} gate is $(1-F_Z)^2$, since it requires that
two encoded qubits be teleported. By using the equation iteratively, the total
probability of gate success can be calculated for a given level of encoding.
These calculations show that, using T$_{3/4}$ teleporters, a four-qubit
encoding is required to achieve a success rate of at least 95\% when performing
a \textsc{cs} gate.

\section{An Incremental Encoding Scheme}
\label{sec:incremental}

%
%
Instead of using a concatenated approach to the encoding, we will add the
component qubits to the encoded state incrementally.  When a $Z$-measurement
error occurs, one of the component qubits is removed from the entanglement, and
this can be corrected by re-encoding using a non-deterministic encoding
circuit. Gates on the overall state can also be performed via re-encoding.

\subsection{Parity Qubit Codes}

%
%
The error-correcting code described earlier (Eq.~\ref{ZMcode}) can be
extended to any number of qubits, in order to protect against multiple
$Z$-measurements. The general form of the code represents
the logical $\Ket{0}$ state as an equal superposition of all states with even
parity (an even number of the component qubits are in the $\Ket{1}$ state), and
the logical $\Ket{1}$ as an equal superposition of all states with odd parity.
This is true whether the method of encoding is incremental or concatenated.  
Hence a general qubit encoded across $n$ component qubits can be written as,
\begin{equation}
\Ket{$\Psi$}^{(n)} = \alpha\Ket{even}^{(n)}+\beta\Ket{odd}^{(n)}.
\end{equation}

%
%
A $Z$-measurement followed by the conditional application of an $X$ gate leaves
the qubit in the correct encoded state, but with one fewer component
qubits than it had initially. After a sufficient number of $Z$-measurements the
encoded qubit will eventually be reduced to the unencoded qubit
$\Ket{$\Psi$}^{(1)} = \alpha\Ket{0}+\beta\Ket{1}$.

\subsection{The Encoder}

\begin{figure}[htb]
\includegraphics[height=80pt,width=180pt]{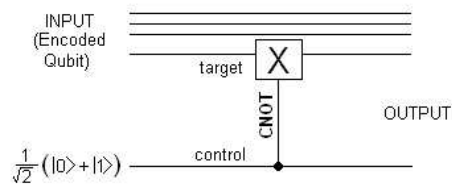}
\caption{The circuit required to increase the number of qubits in
the code state by one.}
\label{Enccir}
\end{figure}

%
%
Instead of doubling the encoding by applying~(\ref{ZMcode}) to each level, we
will  add qubits to the encoding incrementally.  The circuit required to add an
additional qubit to the encoding, consists of a \textsc{cnot} gate and an ancillary qubit
prepared as $(\Ket{0}+\Ket{1})/\sqrt{2}$ (Figure \ref{Enccir}). The ancillary
qubit is the control for the gate, and the target is one of the
components of the encoded qubit. Applying this circuit to a state encoded
across $n$ qubits will produce the correct encoding for the state across $n+1$
qubits.

%
%
By using encoding and teleportation, a non-deterministic gate could be achieved
with arbitrarily high success probability. However, since the encoding circuit
itself involves a non-deterministic gate, we need to teleport this gate and
the success rate of the teleporters must be sufficiently high to produce an
average gain in the encoding.

%
%
For a gate in which one of the inputs is known in advance, that input can be
included in the resource preparation, removing the need for one of the
teleporters and improving the probability of success (the probability of the
gate functioning becomes the same as the probability of the teleportation
succeeding). This is the case for the encoder, since one of the inputs is an
ancillary qubit which will always be in the state $\frac{1}{\sqrt{2}}(\Ket{0}+
\Ket{1})$, and so the encoder functions with the probability of the teleporter
used.

%
%
A failure of the encoder gate results in a $Z$-measurement,
and the encoded state loses a component qubit. Since the encoder is designed
to add one qubit if successful, encoding is equivalent to a
gambling game, with the chances of winning dependent on the
success rate of the encoder. This rate needs to be greater than
50\% to make long-term gains, since we will gain or 
lose a qubit depending on whether the encoder succeeds or fails respectively.



\subsection{Dual-rail vs single-rail}

%
%
The gates described in KLM are all in single-rail form, in that they teleport
only one rail and perform operations solely on it.  If changes other than phase
shifts occur within the gate the number of photons is not conserved, which is a
requirement in a dual-rail system. This limits a dual rail system to performing
\textsc{cs} gates via single-rail teleportation \cite{austin}. However, a
\textsc{cnot} gate is required when creating the incremental encoding or for
performing a encoded \textsc{cnot} efficiently. A teleported \textsc{cs} is not
sufficient, since a Hadamard gate would have to be performed on the individual
qubit beforehand, and doing so takes the overall state out of the code space.
If a teleportation failure occurs, the qubit can no longer be recovered.

\begin{figure}[htb]
\begin{flushleft}
\includegraphics[height=180pt,width=170pt]{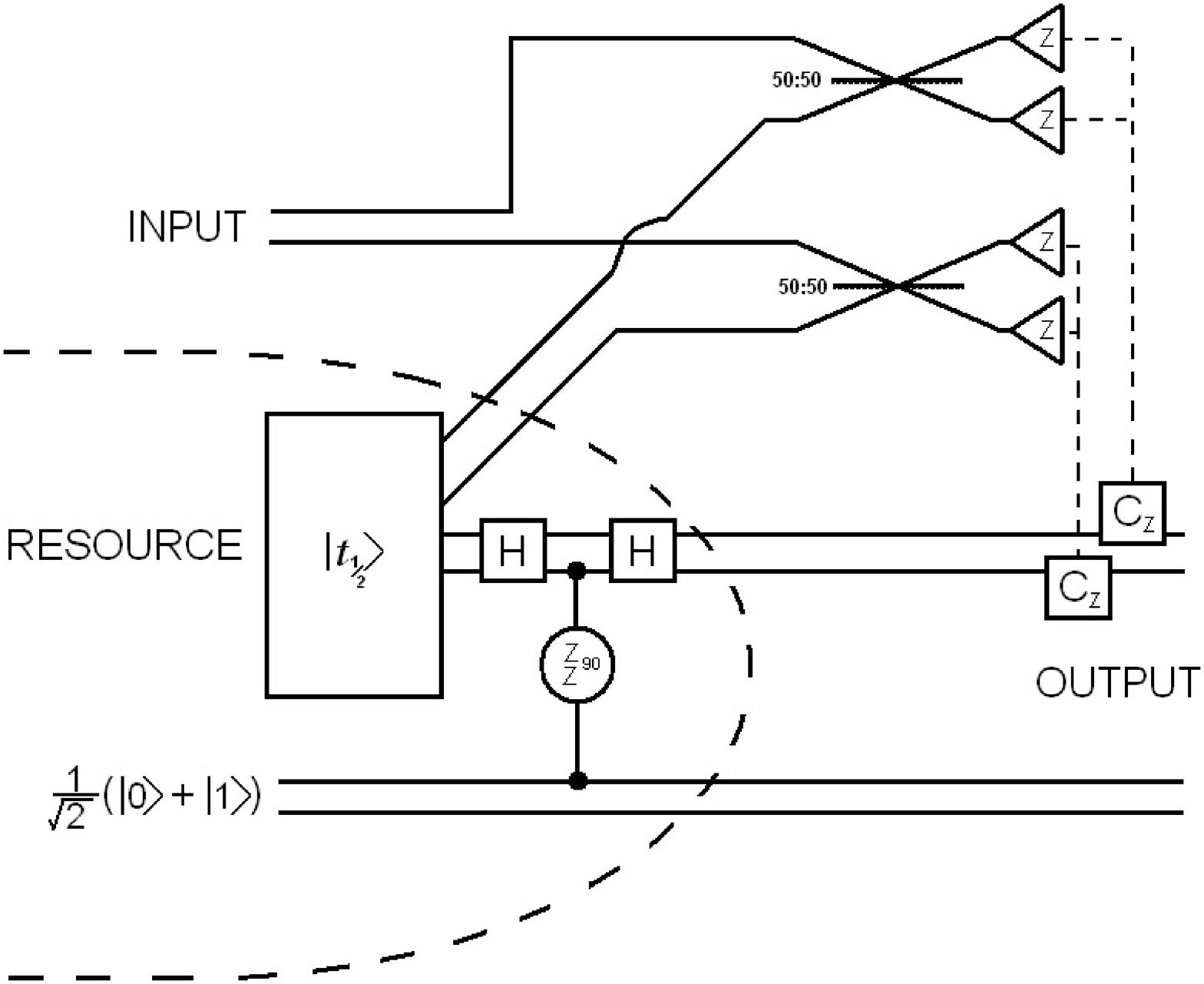}
\end{flushleft}
\caption{The 50\% success dual-rail encoder circuit. C$_Z$
indicates a $Z$-rotation correction dependent on the output
measurements. For higher success rates, a larger entanglement,
\Ket{t$_n$}, is used, and the single beam-splitters are replaced
with arrays of beam splitters that distribute the incoming qubits
equally over $n$ detectors. The circuit shown here is for spatial
encoding, and polarization encoding may vary the design.}
\label{figDRE}
\end{figure}

%
%
A solution is to implement a fully
dual-rail encoder and teleport both rails of the input qubit. It
is known that a dual-rail qubit may be teleported by making a Bell
measurement on the pair of rails \cite{zeil}. In this manner, with
an appropriate resource, the second rail of a physical qubit can
be teleported with certainty if the first rail has been
successfully teleported. This allows the entire gate to be made
dual-rail without reducing the probability of it working. A
dual-rail encoder is not limited to performing only phase
changes. The Hadamards on either side of the
\textsc{cs} can be executed deterministically during the resource
preparation. The disadvantage of this solution is that the size of
the entanglement needed to teleport the input qubit doubles, and
becomes harder to prepare. The teleporter entanglement is a
dual-rail form of the entanglement needed for the single-rail
teleporter. For the T$_{1/2}$ encoder, this is a Bell state. The
circuit for a dual-rail encoder is shown in Figure \ref{figDRE}.

%
%
For a general \textsc{cnot} gate, in which both inputs are unknown, it is
only necessary to use a dual-rail teleporter for the target qubit:
since the control qubit should remain unchanged, a single-rail
teleporter will suffice.

\subsection{Procedures for Implementing Encoded Gates}

%
%
For operations on the encoded states, a series of gates performed on the
component qubits are often required in order to produce a single operation at
the encoded level. This could be achieved using a series of teleported gates
applied to the component qubits; or performing the entire encoded gate at once
by constructing a large entangled state with the gates already applied, and
teleporting every qubit on to it. The KLM proposal uses the latter method.
However, constructing the necessary entangled state requires a large number of
resources. In both cases, re-encoding will also be needed to recover from
teleportation failures.

%
%
If encoding is employed during the computation, it offers new possibilities. By
re-encoding from a subset of the component qubits that have had the desired
operation performed on them, some encoded gates may be performed more
efficiently. This is in general preferable to procedures requiring a series of
encoded gates, as re-encoding would be needed in any case due to teleporter
failures.

\begin{figure}[hHtb]
\begin{flushleft}
\includegraphics[height=130pt,width=170pt]{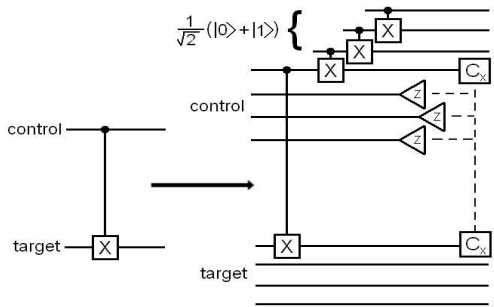}
\end{flushleft}
\caption{The procedure for performing a encoded \textsc{cnot} via
re-encoding. C$_X$ indicates a bit-flip correction dependent on
the outcome of the measurements.} \label{encCNOT}
\end{figure}

%
%
The encoded \textsc{cnot} gate can be made more efficient by using
re-encoding as part of the gate. First a \textsc{cnot} is performed between a
pair of component qubits, one from each encoded qubit, with the logical
control qubit providing the control for this gate (Figure
\ref{encCNOT}). Executing a \textsc{cnot} successfully on one of the
component qubits in the control state produces a subset (of one
qubit) which has undergone the desired rotation. Encoding can now
be performed on the qubit, until the subset reaches the size of
the original state. In the event that teleportation failures cause
the entire subset of component qubits to be lost, the \textsc{cnot} can be
re-attempted on another qubit and the process applied again. Once
the subset reaches the size of the original state, the remaining
unaltered qubits can be measured. If an overall odd parity is
measured, bit-flips must be performed on
both the control and target qubits. This is sufficient to produce
a \textsc{cnot} between the two encoded qubits.

%
%
The $Z_{90}$ rotation will also benefit greatly from being implemented in this
manner. To produce a suitably rotated qubit requires a $Z_{90}$ on one of the
component qubits in the state.  The next step is to encode from the rotated
qubit until the subset again reaches the size of the original state. The
remaining component qubits are then measured. Depending on
the total parity of the measurements, it may be necessary to perform a
$Z_{180}$ rotation to correct the state, as well as the normal bit-flips
required whenever measuring a qubit.

\section{Probability of Encoded Gates}
\label{sec:probs}

%
%
The gates may be attempted until successful, and the problem reduces to one of
maintaining the encoding. The only way in which the total encoded state can be
lost is when the encoder has a sufficiently long run of failures. By adjusting
the level of encoding used and the complexity of the teleporter in the encoder,
the probability of failure can be controlled and kept within an acceptable
limit.

%
%
Each attempt to encode, adds a qubit to the encoding with
probability $p$, which is equal to the probability of the
teleporter succeeding. If the encoder fails, a qubit is removed from
the encoding. What we want to calculate is the probability of
successfully adding a component qubit, without losing the
entire encoded qubit.  This process can be modelled as a random walk on a discrete,
one-dimensional lattice, with probabilities of moving in the
positive or negative directions of $p$ and $(1-p)$ respectively.
The lattice has absorbing boundaries at $R$ and $L$ ($R>L$). These
boundaries correspond to successfully adding a qubit, and to losing
\emph{all} the component qubits in the encoding.

%
%
Starting from location $m$ on the lattice, $L\le m\le R$, the
probability $P_{R,m}$ of reaching $R$ without first reaching $L$
is given by the solution to the equation \cite{vankamp},
\begin{equation}
    P_{R,m} =  pP_{R,m+1}+(1-p)P_{R,m-1}.
\end{equation}
This is a homogeneous second order difference equation and can be
solved simply by trial solutions yielding,
\begin{equation}
    P_{R,m}=\left\{ \begin{array}{ll}
         \frac{m-L}{R-L} & \mbox{if } p = \frac{1}{2}\\
        \frac{1-\beta^{m-L}}{1-\beta^{R-L}}
         & \mbox{if } p \not= \frac{1}{2}\end{array} \right.,
\end{equation}
where $\beta = (1-p)/p$. The probability of successfully adding a
qubit without losing the encoding is therefore
$P_\mathrm{add}\equiv P_{R,0}$ with $L=-w$ and $R=1$, where $w$ is
the current number of components making up the encoded qubit, that is:
\begin{equation}
    P_\mathrm{add}=\left\{ \begin{array}{ll}
         \frac{w}{w+1} & \mbox{if } p = \frac{1}{2}\\
        \frac{1-\beta^{w}}{1-\beta^{w+1}}
         & \mbox{if } p \not= \frac{1}{2}\end{array} \right..
\end{equation}

\begin{figure}[htb]
\begin{center}
  \begin{tabular}{cc}
   \includegraphics[scale=.5]{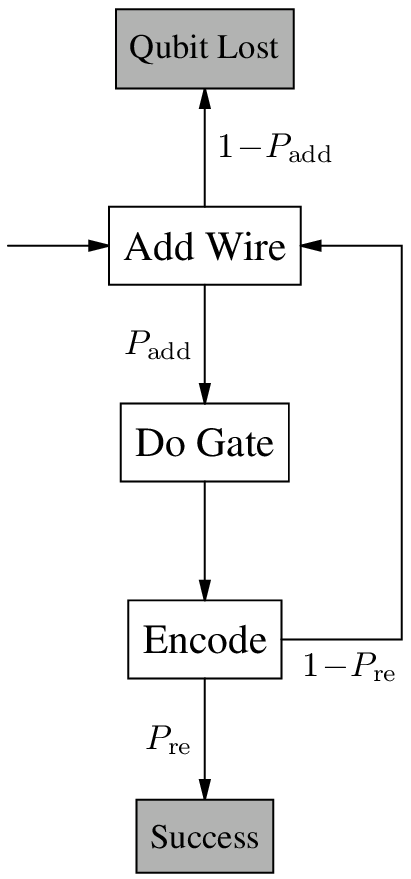}\;\;\;  & \;\;\;
\includegraphics[scale=.5]{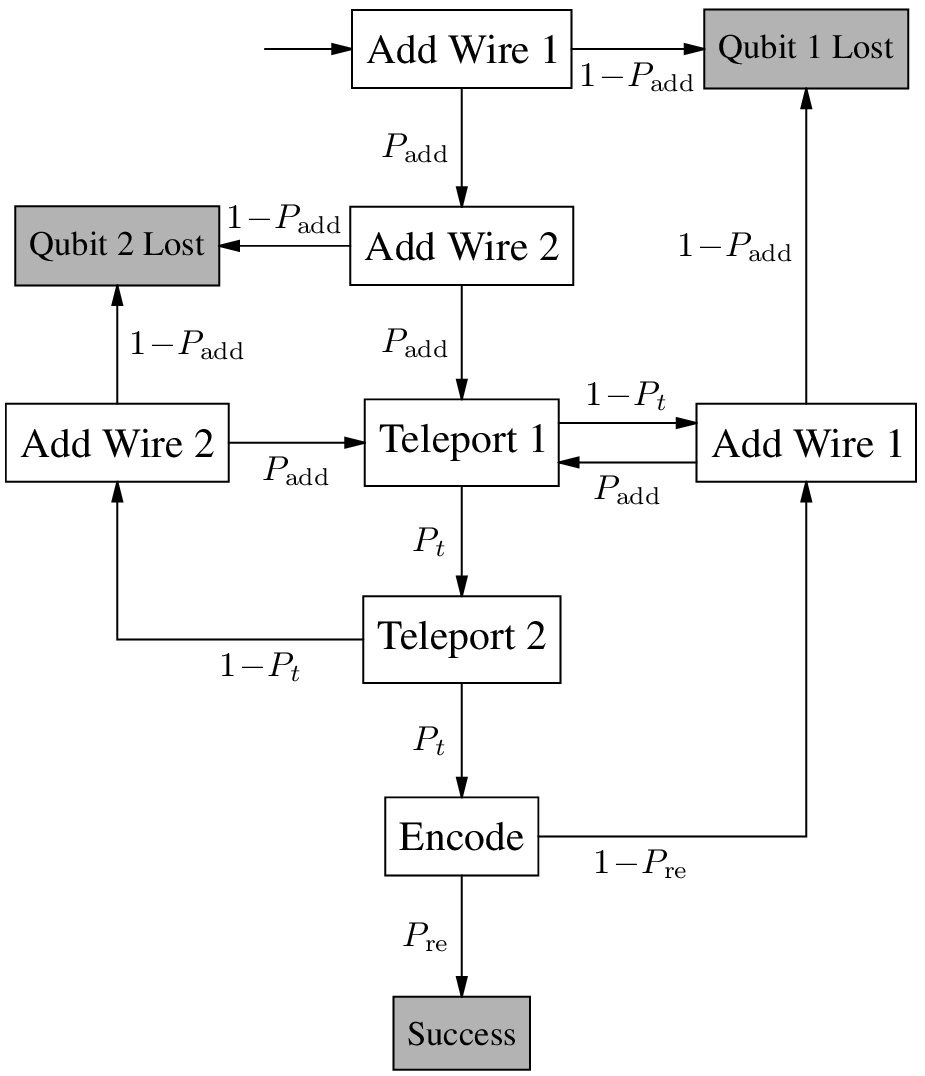} \\
(a) & (b)
   \end{tabular}
\end{center}
\caption{Algorithm for performing gates on the encoded qubits.
  (a) a single qubit $Z_{90}$ gate and, (b) a \textsc{cnot} gate.  The various
  probabilities are explained in the text.}
 \label{flowcharts}
\end{figure}

%
%
Algorithms for performing gates on the encoded qubits are given in
Figure~\ref{flowcharts}. There are essentially three stages to the
algorithm: adding a qubit to the existing encoding, performing the
gate on the new qubit, and re-encoding from the new qubit (re-building the entire encoded
qubit from the new component).

%
%
Consider first a $Z_{90}$ gate operating on a single encoded qubit.  Following the
algorithm in Figure~\ref{flowcharts}(a), the first step is to add
a qubit to the encoding in order to avoid risking the existing
entanglement, this occurs with probability $P_\mathrm{add}$ as
discussed above. Next we perform the gate on the new qubit, this
is deterministic. Finally, we re-encode from the new qubit. The
probability of re-encoding, $P_\mathrm{re}$, can be calculated
from the same random walk argument as before starting from $m=1$,
only now the absorbing boundaries are at $L=0$ (loss of a qubit) and
$R=w$ (successful re-encoding to $w$ qubits). Hence we have
\begin{equation}
    P_\mathrm{re}=\left\{ \begin{array}{ll}
         \frac{1}{w} & \mbox{if $p = \frac{1}{2}$}\\
        \frac{1-\beta}{1-\beta^w}
         & \mbox{if $p \not= \frac{1}{2}$}\end{array} \right..
\end{equation}

%
%
The probability of successfully applying the gate and re-encoding
is $P_\mathrm{re}$, so on average we will need $1/P_\mathrm{re}$
attempts. Each unsuccessful attempt destroys the qubit we added, so
we will have to add a new qubit with each new attempt. This gives
an overall probability of success for the algorithm for single
encoded-qubit gates of $(P_\mathrm{add})^{1/P_\mathrm{re}}$.

%
%
For a \textsc{cnot} gate, the algorithm is only slightly more complicated
(Figure~\ref{flowcharts}(b)). First we add an extra component qubit to both
the encoded target and control qubits.  Then we attempt two
teleportations with appropriate resource for a \textsc{cnot}.  If either of
these are unsuccessful, we will lose only a single qubit, which we
will have to add again before attempting both teleportations.  If
both are successful, we need only re-encode from the target qubit
as the control qubit will not be affected by the action of the
\textsc{cnot}. Successfully performing both teleportations (each with
probability $P_t$) and re-encoding, will happen with probability
$P_t^2P_\mathrm{re}$. Note that the teleporters used for the \textsc{cnot}
(probability $P_t$) need not be the same as the teleporters used
for the encoder (probability $p$). Following the previous argument
and taking into account the initial addition of a qubit gives an
overall probability of success of
$(P_\mathrm{add})^{1+1/(P_t^2P_\mathrm{re})}$.

%
%
For simplicity we shall take the probability of successfully
performing a gate on the encoded qubits $P_\mathrm{enc}$ as the
probability of performing a \textsc{cnot} as this is the lower probability.
That is, $P_\mathrm{enc}=(P_\mathrm{add})^{1+1/(P_t^2P_\mathrm{re})}$.
These results can then be used to examine how the probability of
the encoded gate functions as we use larger encoded states.  By
requiring a fixed total probability of performing an entire
computation, an expression can be found for the number of gates
that can be performed given an encoding of size $w$ of the qubits.
The probability of an entire computation succeeding is given by
the product of the success probabilities of each gate in the
circuit.  Hence the total probability for a computation containing
$n$ gates succeeding is given by
$P_\mathrm{tot}=P_\mathrm{enc}^n$. This assumes that these are all
\textsc{cnot} gates, and places a \emph{lower} bound on the number of
non-deterministic gates that can be performed.

In general,
\begin{equation}
  \label{eq:scale}
  n = \frac{\log(P_\mathrm{tot})}{\left(1+\frac{1}{P_t^2P_\mathrm{re}(w)}\right)\log P_\mathrm{add}(w)}.
\end{equation}
When the T$_{1/2}$ encoder is used, the expression for $n$
becomes:
\begin{equation}\label{eq:n}
n =
\frac{\log(P_\mathrm{tot})}{\left(\frac{w}{P_t^2}+1\right)\left(\log(w)-\log(w+1)\right)}.
\end{equation}
As $w$ increases, the magnitude of the denominator in~(\ref{eq:n}) will also increase. From this, it is clear that the number of
gates that can be performed with a set probability actually
decreases as the number of component qubits increases. Therefore the
T$_{1/2}$ encoder cannot be used to protect the encoded qubits.

\begin{figure}[htbH]
\begin{center}
\includegraphics[scale=.9]{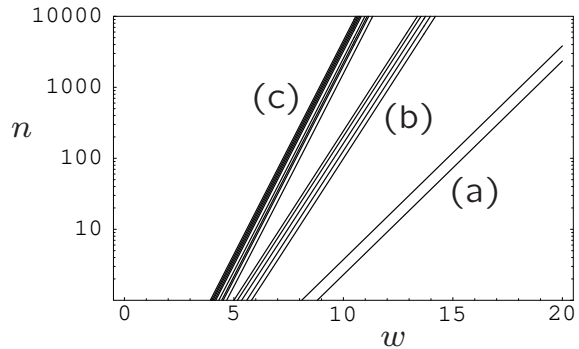} 
\end{center}
\caption{Number of component qubits ($w$) needed in order to perform $n$ encoded \textsc{cnot}
gates
  with a total probability of 99\%. (a) Using $T_{2/3}$ encoder for
  the initial qubit and re-encoding, and $T_{1/2}$ and $T_{2/3}$ for the
  \textsc{cnot}; (b) using $T_{3/4}$ encoder for initial qubit, and
  $\{T_{3/4}$,\dots,$T_{2/3}\}$ for re-encoder, and
  $\{T_{3/4}$,\dots,$T_{1/2}\}$ for the \textsc{cnot}; (c) using $T_{4/5}$ encoder
  for initial qubit, and $\{T_{4/5}$,\dots,$T_{2/3}\}$ for re-encoder, and
  $\{T_{4/5}$,\dots,$T_{1/2}\}$ for the \textsc{cnot}.} \label{figscale}
\end{figure}

%
%
For teleporters with probability higher than $1/2$, the
relationship is shown in Figure \ref{figscale}. The results show
that there is relatively little benefit in using more complicated
teleporters for the re-encoding or the \textsc{cnot} gate. This suggests
that for these steps we use the $T_{2/3}$ and $T_{1/2}$
teleporters respectively. The success probabilities for a given
number of component qubits found here are very similar to those for the
encoded \textsc{cs} gate in the original KLM proposal.

\section{Resources for an Encoded  Gate}
\label{sec:resources}

%
%
In order to estimate the resource usage of our algorithm we will
first calculate the mean number of times that we use the encoders
and teleporters of each stage in the algorithm in order to perform
a successful encoded gate. After this, we will break down the
resources into more fundamental units.

%
%
The key to calculating the number of gates used is the mean
first-passage times in the random walk model. This provides the
average number of steps that occurs before reaching a lattice
boundary, which in turn corresponds to success or failure of the qubit-adding
stage, or of the re-encoding stage of the algorithm.

%
%
Using the previous random walk model with absorbing boundaries,
the mean number of steps to the $R$ boundary ($\ave{N_{R,m}}$)
without hitting the $L$ boundary, can be calculated as a solution
the following difference equation \cite{vankamp},
\begin{multline}
  \label{eq:first-pass}
  P_{R,m}(\ave{N_{R,m}}-1) =  pP_{R,m+1}\ave{N_{R,m+1}}\\+(1-p)P_{R,m-1}\ave{N_{R,m}}
\end{multline}
which again can be solved by trial solutions. For $p>1/2$ the
solution is,
\begin{multline}
\ave{N_{R,m}} = [ R(\beta^L-\beta^m)(\beta^L+\beta^R)\\
-m(\beta^L+\beta^m)(\beta^L-\beta^R) \\
+2L(\beta^{L+m}-\beta^{L+R})]\\/[(2p-1)(\beta^L-\beta^m)(\beta^L-\beta^R)].
\label{eq:first-passage}
\end{multline}

For the addition stage, in which one qubit is to be added to the
encoded qubit which has $w$ component qubits, the mean
first-passage time for successfully adding the qubit is
$\ave{N_\mathrm{add}}=\ave{N_{R,0}}$ with $L=-w$ and $R=1$:
\begin{equation}
\ave{N_\mathrm{add}}=
                \frac{(1-\beta^{w})(1+\beta^{w+1})-2w\beta^{w}(1-\beta)}
        {(2p-1)(1-\beta^{w})(1-\beta^{w+1})}
\label{eq:tS}
\end{equation}
This is the mean number of times we will use the encoder for each
pass through the qubit-adding stage of the algorithm without losing 
the entire encoded qubit.

%
%
As the number of component qubits tends to infinity, these values will approach
a limit, given by the inverse of the expectation value for the encoder gain.
For example, the $T_{2/3}$ encoder has an expectation value of $2/3\times
1-1/3\times 1=1/3$ qubits per attempt. This gives an average of 3 attempts
needed to add one qubit when the boundary effects are negligible. For a finite
number of qubits, the average is less, since the majority of the outcomes that
succeeded after a long sequence of attempts now result in the loss of the
entire encoded qubit.  Practically, the number of component qubits only needs
to be of the order of ten for the $T_{2/3}$ and five for higher teleporters,
for these limits to be a good approximation.

The mean number of uses of the encoder in each successful attempt
to re-encode will be $\ave{N_\mathrm{re}}=\ave{N_{R,1}}$ with
$L=0$ and $R=w$:
\begin{equation}
\ave{N_\mathrm{re}}=
                \frac{w(1-\beta)(1+\beta^{w})-(1+\beta)(1-\beta^w)}
        {(2p-1)(1-\beta)(1-\beta^{w})}.
\label{eq:tI}
\end{equation}
The limiting behaviour of $\ave{N_\mathrm{re}}$ with increasing
$w$ goes as $w/(2p-1)$ and so is linear in $w$.

Similarly, the mean number of uses of the encoder in each
unsuccessful attempt to re-encode, $\ave{N_\mathrm{fre}}$, can
also be obtained from equation~(\ref{eq:first-passage}) for
$\ave{N_{R,-1}}$ with $L=-w$, $R=0$ and setting
$p\rightarrow(1-p)$ (swapping left and right directions):
\begin{equation}
\ave{N_\mathrm{fre}}=
                \frac{(\beta^w+\beta)(\beta^w-1)-2w(\beta^{w+1}-\beta^w)}
        {(2p-1)(\beta^w-\beta)(1-\beta^{w})}.
\label{eq:tF}
\end{equation}

%
%
We can now calculate the total mean number of times we use the
encoders and teleporters of each stage for a successful encoded
gate. The average number of times we use the encoder to add a qubit
is $2\ave{N_\mathrm{add}}$ for the initial attempts, plus
$\ave{N_\mathrm{add}}$ for all but the last successful attempt to
perform the teleporters and re-encoding:
\begin{equation}
  \label{eq:E+1}
  \ave{E_\mathrm{add}}=\left(\frac{1}{P_t^2P_\mathrm{re}}+1\right)\ave{N_\mathrm{add}}.
\end{equation}
The average number of applications of the re-encoding is
$\ave{N_\mathrm{re}}$ plus $\ave{N_\mathrm{fre}}$ for all the
failed attempts:
\begin{equation}
  \label{eq:Ere}
  \ave{E_{re}}= \left(\frac{1}{P_\mathrm{re}}-1\right)\ave{N_\mathrm{fre}}+\ave{N_\mathrm{re}}
\end{equation}
and finally, the average number of applications of the teleporters in gate
stage is
\begin{equation}
  \label{eq:Tg}
  \ave{T_g} = \frac{1}{P_t^2P_\mathrm{re}}.
\end{equation}

We can choose a different teleporter for each of the stages: qubit
addition, re-encoding, and doing the actual gate, and we will
characterise these three teleporters by the numbers $n_a$,$n_r$
and $n_t$ respectively, where the probability of the teleporter is
given by $n/(n-1)$. This allows us to calculate the average number 
of resources needed to perform an encoded gate. For instance, a 
\textsc{cnot} on two qubits, each encoded across four component qubits, with 
a 95\% probability of success would require on average 7.5 
$T_{1/2}$ \textsc{cnot} gates, 16.0 $T_{3/4}$ encoders and 5.7 $T_{2/3}$ 
encoders.

%
%
Since our proposal uses a slightly different set of circuits from
those in KLM, a comparison of the resources used in each is
difficult. However, by assuming a standard method for constructing
these circuits, some comparison can be made. The entanglements 
required by the teleporters can be produced relatively efficiently 
using the KLM \textsc{cs} circuit and circuits which eliminate certain 
component states from the overall state of the resource qubits. 
These elimination circuits are described in Appendix~\ref{sec:elim}. 
A method for constructing the family of teleporters used by KLM is 
detailed in Appendix~\ref{sec:cnstrtn}.

For the purposes of this comparison, we ignore the degree to which
these resources could be constructed in parallel, and simply
consider the minimum number of circuits that could produce the
required resources. The resources compared are those needed on
average to implement a gate on two qubits, each encoded
across four component qubits, with a 95\% probability of success.

First we consider our scheme. Each use of an encoder will require 
$2n$ physical \textsc{cs}'s and $n-1$ elimination circuits. An encoded \textsc{cnot} 
gate will consume $n^2+n$ physical \textsc{cs}'s and $2(n-1)$ elimination 
circuits.  Hence, in order to implement an encoded \textsc{cnot} gate the 
minimum number of physical \textsc{cs}'s that will be needed will be
\begin{equation}
    \ave{N_{CS}} = 2n_a \ave{E_\mathrm{add}} + 2n_r
\ave{E_{re}}+(n_t^2+n_t)\ave{T_g} \label{eq:Ncs}
\end{equation}
and the minimum number of elimination circuits will be
\begin{equation}
    \ave{N_{elim}} = (n_a-1) \ave{E_\mathrm{add}} + (n_r-1)
\ave{E_{re}}+2(n_t-1)\ave{T_g} \label{eq:Nelim}
\end{equation}
Given a desired probability of operation, we can easily
numerically solve the equation for $P_\mathrm{enc}$, the
probability of success of the gate, for $w$ which we can then use
to calculate $N_{CS}$ and $N_{elim}$. For a 95\% probability encoded 
\textsc{cnot}, around 90 physical \textsc{cs}'s and 32 elimination circuits are 
required.

In the original KLM proposal the resource usage was estimated by considering
the average number of teleportations that must be attempted in order to
successfully perform an encoded gate. The supporting paper for KLM \cite{thresholds} states
that, using $T_{3/4}$ and logical states encoded across four component qubits,
the expected number of teleported \textsc{cs} needed to perform an encoded
\textsc{cs} is less than 250.  If the teleportation entanglements are assumed
to be produced by a method of elimination, each teleported \textsc{cs} would
require four elimination circuits and nine \textsc{cs} circuits. From this, it
can be seen that the expected requirements for an encoded gate would be less
than 1000 elimination circuits and less than 2250 \textsc{cs} circuits.  This
indicates that our methods would have a significant advantage over the original
KLM techniques in resource usage, while achieving equal probabilities of
success when performing equivalent gates.

\begin{figure}[tbh]
\begin{center}
\begin{tabular}{cc}
\includegraphics[scale=.5]{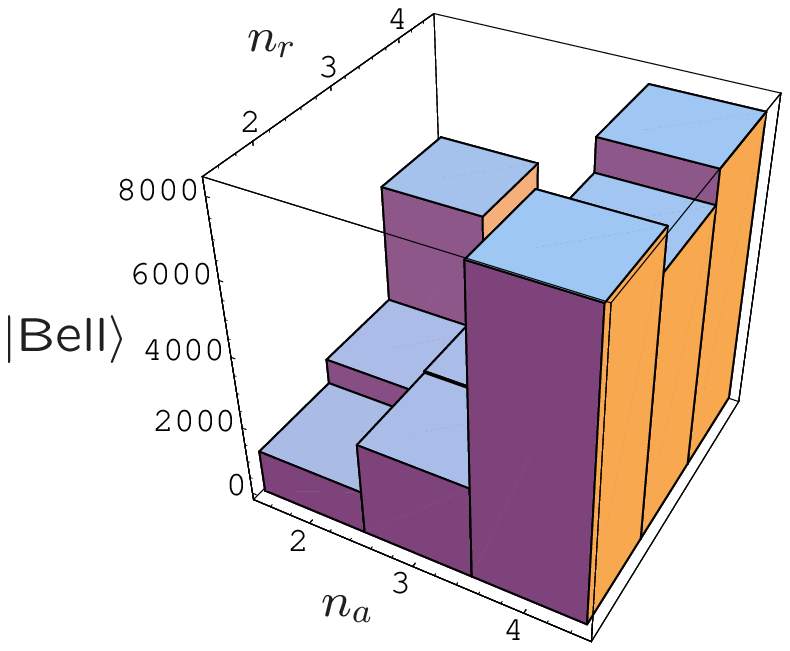} &
\includegraphics[scale=.5]{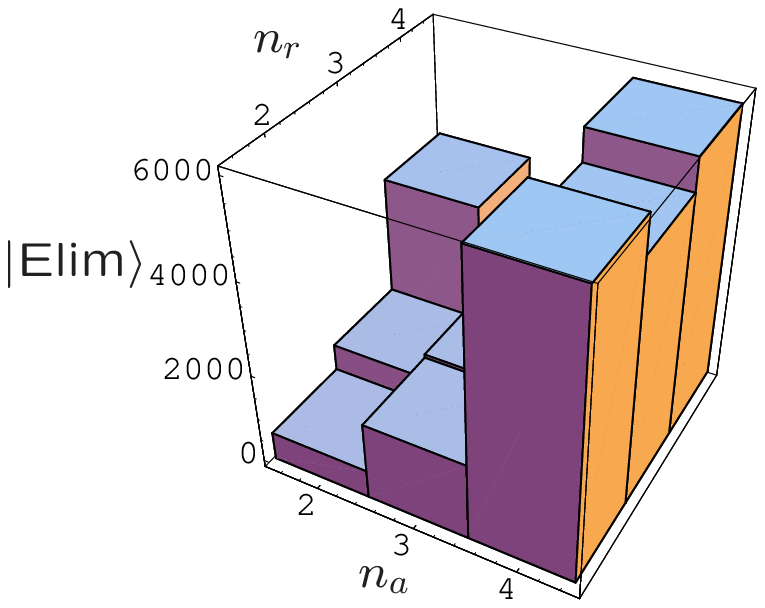}\\
(a) & (b)
\end{tabular}
\end{center}
\caption{Resource requirements to perform an encoded \textsc{cnot} with
probability 95\% as a function of the teleporters used. In both
cases the gate teleporters were $T_{1/2}$. The adding-a-component 
stage used a $T_{n_a/(n_a+1)}$ teleporter, and the re-encoding stage used
$T_{n_r/(n_r+1)}$ teleporter. (a) Number of Bell states used. 
(b) Number of elimination states used.} \label{fig:paraRes}
\end{figure}

To better analyse the resources used in performing an encoded gate and to take
into account possible parallel resource construction, we consider that we have
available at our disposal massively parallel state factories that can produce
Bell states ($\ket{\textrm{Bell}}=(\ket{01}+\ket{10})/\sqrt{2}$), and
elimination states
($\ket{\textrm{Elim}}=\bigl(\ket{10001011}+\ket{01101001}+\ket{01110100}\bigr)/\sqrt{3}$).
A way of producing the elimination states is given in Appendix~\ref{sec:elim}.
With these two states as resources, the entanglement required for the encoder
can be built as outlined in Appendix~\ref{sec:cnstrtn}. We will assume that in
the event of a teleportation failure, the entire resource is discarded and so
we will end up discarding a large fraction of the states produced until we
perform the equivalent of around 90 physical \textsc{cs}'s and 32 elimination
circuits as described earlier.  In practice, it is likely that part of the
entanglement could be reused.  How many of the resource states would we consume
to perform a 95\% probability encoded \textsc{cs} gate? The results are shown
in Figure~\ref{fig:paraRes}, and yield a minimum of approximately 1300 Bell
states and 620 elimination states.  Of course there is no recycling used in the
above estimate and this would reduce these figures. The results also indicate
that the is no advantage to using more sophisticated teleporters than $T_{2/3}$
in any of the stages of our procedure.

\section{Conclusion}

In this paper, the concepts and techniques put forward in KLM were reworked
using a incremental form of encoding against $Z$-measurement errors rather than
using a concatenated approach. The general encoding circuit for producing this code
was described, and can be used to implement encoded gates with improved
efficiency.  This allows computations
to be safely performed with fewer resources than would be required by the
original KLM scheme.

The random walk model for the encoding process was used to verify that the
proposal scales efficiently for teleporter probabilities greater
than 0.5. The scaling allows for a large number of gates to be
performed safely in the ideal case using a relatively low level of
encoding. Although the scaling is still of the same order as that
achieved by the original proposal, the improved efficiency makes
this scheme preferable for implementation, if the resources can be
produced reliably.

Recently, the cluster state models have shown great promise for
efficiently implementing LOQC. Due to the similarity between these
processes and those used in our scheme, it is likely that some of
the techniques described here can also be applied to the
development of cluster state proposals and vice-versa.

\begin{acknowledgments}
We acknowledge the support of the Australian Research Council, the ARO and the
Queensland State Government. AG also acknowledges support by the New Zealand
Foundation for Research, Science and Technology under grant UQSL0001.  We would
like to thank Michael Nielsen for discussions and suggesting this project, and
Andrew White, Geoff Pryde, Jeremy O'Brien, Nathan Langford and Nicole Muir for
helpful comments on this manuscript.
\end{acknowledgments}

\appendix
\section{Elimination Circuits}
\label{sec:elim}

Here we discuss the use of elimination circuits for producing the entanglement
resource required for the higher order teleporters. The construction is related
to the \textsc{cs} gate, but instead of inserting a sign change on a particular
state component, it eliminates that component. Consider preparing
the entanglement resource for the single rail $T_{2/3}$ teleporter. We start
with the two dual rail single photon states $\ket{10}$ and $\ket{01}$. The dual
rails of each of the photons are mixed on a $1/7$ reflectivity beam-splitter
giving:
\begin{eqnarray}
\ket{10}&\rightarrow&
\frac{6}{\sqrt{42}}\ket{10}+\frac{1}{\sqrt{7}}\ket{01}\\
\ket{01}&\rightarrow&\frac{1}{\sqrt{7}}\ket{10}-\frac{6}{\sqrt{42}}\ket{01}
\end{eqnarray}
Now include two single photon Fock states in modes 2 and 5:
\begin{multline}
\ket{\text{input}}=\bigl(
\frac{6}{\sqrt{42}}\ket{110}+\frac{1}{\sqrt{7}}\ket{011}
\bigr)_{123}\otimes\\\bigl(
\frac{1}{\sqrt{7}}\ket{110}-\frac{6}{\sqrt{42}}\ket{011}\bigr)_{456}
\end{multline}
Modes 3 and 4 are mixed on a $1/2$ reflectivity beam-splitter.
\begin{multline}
\frac{1}{7}\bigl(\sqrt{3}\ket{010111}-3\sqrt{2}\ket{010210}+
\sqrt{3}\ket{011011}\\
+3\sqrt{2}\ket{012010}+\ket{110011}-\sqrt{3}\ket{110110}\\
+\sqrt{3}\ket{111010}\bigr)_{123456}
\end{multline}
Next, modes 2 and 3, and modes 4 and 5, are mixed on $1/3$
reflectivity beam-splitters. The conditional state obtained when
one and only one photon is found in mode 2 and simultaneously one
and only one photon is found in mode 5 is given by:
\begin{multline}
\frac{\sqrt{2}}{21}\bigl(
\ket{0011}+\ket{0101}+\sqrt{2}\ket{1001}-\\\ket{1010}+\ket{1100}
\bigr)_{1346}
\end{multline}
The effect of this conditional step is to remove the terms
containing photon pairs. Modes 3 and 4 are then mixed on another
$1/2$ reflectivity beam-splitter and we get
\begin{equation}
\frac{2}{21}\bigl(\ket{0011}+\ket{1001}+\ket{1100}\bigr)_{1346}
\end{equation}
which is the required entangled state, produced with a probability of
$\frac{12}{441}\approx 0.027$.

The entanglement resource for a \emph{dual} rail $T_{2/3}$ teleporter can be
produced in a similar way. We first need to produce the states
\begin{gather}
\frac{6}{\sqrt{42}}\ket{0110}+\frac{1}{\sqrt{7}}\ket{1001}\\
\frac{1}{\sqrt{7}}\ket{1001}-\frac{6}{\sqrt{42}}\ket{0110}
\end{gather}
These can be produced from separable inputs via some linear optics elements and
two \textsc{cs} gates. As before we include two single photon Fock states in
modes 2 and 5:
\begin{multline}
\ket{\text{input}}=\bigl(
\frac{6}{\sqrt{42}}\ket{01110}+\frac{1}{\sqrt{7}}\ket{10011}
\bigr)_{ab123}\otimes\\\bigl(
\frac{1}{\sqrt{7}}\ket{01110}-\frac{6}{\sqrt{42}}\ket{10011}\bigr)_{cd456}
\end{multline}
The protocol now proceeds just as before on modes 2, 3, 4 and 5. The
resultant state is then
\begin{equation}
\frac{2}{21}\bigl(\ket{10001011}+\ket{01101001}+\ket{01110100}\bigr)_{ab13cd46}
\end{equation}
which is the required entangled state for dual rail, produced
again with a probability of $\frac{12}{441}\approx 0.027$.

\section{Constructing Teleportation Entanglement}
\label{sec:cnstrtn}

The general formula for the teleporter entanglement \Ket{t$_n$} is given in
Eq.~(\ref{tnform1}). To produce the entanglement \Ket{t$_{n+1}$}, first assume
that the entanglement \Ket{t$_n$} can be made:
\begin{multline}
\Ket{t$_n$}=\frac{1}{\sqrt{n+1}}(\Ket{0}_C\Ket{0}^{n-1}\Ket{1}^n
    \\+\Ket{1}_C\sum_{j=1}^n\Ket{1}^{j-1}\Ket{0}^{n-j} \Ket{0}^j
    \Ket{1}^{n-j}).
\end{multline}
Now add a Bell state of the form $\frac{1}{\sqrt{2}}(\Ket{01}+\Ket{10})_{AB}$,
\begin{multline}
\Ket{combined}=\frac{1}{\sqrt{n+2}}(\Ket{0}_A\Ket{0}_C\Ket{0}^{n-1}
    \Ket{1}_B\Ket{1}^n\\+\Ket{1}_A\Ket{0}_C\Ket{0}^{n-1}
    \Ket{0}_B\Ket{1}^n\\
    +\Ket{1}_A\Ket{1}_C
         \sum_{j=1}^n\Ket{1}^{j-1}\Ket{0}^{n-j}
    \Ket{0}^j\Ket{0}_B\Ket{1}^{n-j})
\end{multline}
then it can be linked to the \Ket{t$_n$} state via an elimination circuit,
producing \Ket{t$_{n+1}$}.  The elimination operation is performed on qubits
$B$ and $C$, removing any states containing $\Ket{11}_{BC}$.  That is,
\begin{align}
\Ket{combined}\rightarrow&\frac{1}{\sqrt{n+2}}(\Ket{0}^{n+1}\Ket{1}^{n+1}
+\Ket{1}\Ket{0}^{n+1}\Ket{1}^n\nonumber\\
    &\;\;\;+\sum_{j=1}^{n}
    \Ket{1}^{j+1}\Ket{0}^{n-j}\Ket{0}^{j+1}\Ket{1}^{n-j})\\
         =&\sum_{j=0}^{n+1}\Ket{1}^j\Ket{0}^{(n+1)-j}\Ket{0}^j\Ket{1}^{(n+1)-j}
\\=& \Ket{t$_{n+1}$}
\end{align}

For $n=1$, the required entanglement is simply a Bell state:
$\frac{1}{\sqrt{2}}(\Ket{01}+\Ket{10})$. If a single-rail
teleporter is to be used, these states can be created
deterministically using a beam splitter. In dual-rail, they
require a \textsc{cs} gate in order to produce them.

From this, it can be seen that preparing the state $\Ket{t$_n$}$
requires $n-1$ successfully performed elimination circuits. If the
teleporter is to be dual-rail, an additional $n$ \textsc{cs} gates are
also required.

In building up the entanglement, it is more efficient to perform
each \textsc{cs} and elimination circuit separately, then link them
together using teleportation. Basic 50\% success teleporters can
be used, as these require no extra resources. The inputs to the
circuits are single photons in the state
$\frac{1}{\sqrt{2}}(\ket{01}+\ket{10})$, which is easily produced
via a beam-splitter. For the elimination circuit, these states can
be weighted, removing the need to apply weightings later.



\end{document}